\documentclass[prl,aps,amssymb,twocolumn,showpacs,superscriptaddress,preprintnumbers]{revtex4}
\usepackage{graphicx}

\begin{document}


\title{Universal output directionality of ultrahigh-$Q$ single modes in a deformed microcavity with low index of refraction}

\author{Sang-Bum Lee}
\address{School of Physics and Astronomy, Seoul National University, Seoul 151-747, Korea}
\author{Jeong-Bo Shim}
\address{Department of Physics, Korea Advanced Institute of
Science and Technology, Daejon 305-701, Korea}
\author{Sang Wook Kim}
\address{Department of Physics Education, Pusan National University, Busan 609-735, Korea}
\author{Juhee Yang}
\address{School of Physics and Astronomy, Seoul National University, Seoul 151-747, Korea}
\author{Songky Moon}
\address{School of Physics and Astronomy, Seoul National University, Seoul 151-747, Korea}
\author{Jai-Hyung Lee}
\address{School of Physics and Astronomy, Seoul National University, Seoul 151-747, Korea}
\author{Hai-Woong Lee}
\address{Department of Physics, Korea Advanced Institute of Science and Technology, Daejon 305-701, Korea}
\author{Kyungwon An}
\email{kwan@phya.snu.ac.kr}
\address{School of Physics and Astronomy, Seoul National University, Seoul 151-747, Korea}

\date{\today}

\begin{abstract}
Experimental investigation of the characteristics of quasi-bound states of a quadrupole deformed microcavity has revealed five distinct mode groups in cavity emission spectra with cavity quality factors different by orders of magnitude and consistently with much different intracavity mode distributions but with almost universal far-field emission patterns. These universal directionality of high $Q$ modes are explained by a subtle manifestation of unstable manifolds of classical chaos in the formation of quasi-bound states.
\end{abstract}


\maketitle

How classical trajectory of a particle is associated with its quantum wave function \cite{Q-chaos} is a frequently encountered theme in various branches of modern physics, ranging from Anderson localization \cite{Anderson} in a disordered conductor and its optical version in random media \cite{Lagendijk,Genack} to electronic wave functions in a high external field in hydrogen \cite{hydrogen} or in a quantum dot \cite{quantumdot} and microwave distribution in a Sinai billiard \cite{Sridhar}. Closely related to this theme, understanding mode formation in microscopic cavities is an important task from a practical point of view, especially in designing efficient laser cavities for photonics applications \cite{photonics}. Particularly, mode formation in deformed microcavities of stadium or quadrupole shape has drawn much attention since they not only support collimated directional output \cite{Noeckel} desirable for efficient microlasers but also provide a testing ground for classical and quantum (wave) chaos \cite{Reichl92}.

The output directionality observed from deformed microcavities has been explained either by ray dynamics based on chaos theory \cite{Lichtenberg83, Noeckel}, or by the nature of modes obtained from Maxwell's wave equations. Adequacy of each approach depends on several factors, but most importantly on the size of microcavity with respect to the wavelength of interest or the size parameter $nkr$ where $n$ is the refractive index of the cavity medium, $k=2\pi/\lambda$ the wave vector with $\lambda$ the wavelength and $r$ a representative radius of the cavity.

Most studies on output directionality have been performed on relatively large cavities with $nkr\sim 10^3$ or so \cite{Rex02,Gmachl02,Harayama03,Schwefel04}, for which analysis of ray dynamics in phase space, corresponding to classical limit of $nkr\rightarrow \infty$, has been successful in correctly predicting the output emission properties of stadium, quadrupole or hexadecapole microcavities \cite{Schwefel04}.

Although modes of cavity are inevitably involved in those experiments with large $nkr$, measuring actual mode spectra and resolving individual modes have not been reported mainly due to the fact that the free spectral range (FSR) of a mode in such large cavities is not much larger than the linewidths of individual modes of relatively low cavity quality factor $Q$, associated with the observed output directionality.

There have been experiments performed on comparably smaller quardrupole-deformed microcavities (QDM's) with $nkr\sim 200$ \cite{Lee02}, corresponding to an interesting regime where wave and particle nature of light may coexist. Recently scar modes, {\em i.e.} quantum localization or a quasi-bound state along an unstable periodic orbit (UPO) in ray picture have been observed in deformed microcavities with $nkr\sim 10^2 - 10^3$ \cite{Lee02,Rex02,Gmachl02,Harayama03}. However, spectra of high-$Q$ scar modes with well-defined FSR have been only observed in QDM's of low index of refraction ($n=1.36$) with $nkr\sim 200$ \cite{Lee02} due to the aforementioned technical reasons.

The aim of the present work is to elucidate the interplay between wave and particle natures of light in mode formation in a QDM of properly chosen size parameter ($nkr\sim 200$) with direct comparison between experiment and theory with the same size parameter. In particular, we report observation of five different well-defined mode groups including scar modes, with $Q$ values ranging from $10^4$ to $10^6$, in both spectrum and output directionality in a QDM with a large deformation to enable complete classical chaos for the ray undergoing total internal reflections. Interestingly, these modes exhibit similar output directionality regardless of their different spatial distributions inside the cavity. This universal directionality of high $Q$ modes can be explained by an unexpected manifestation of ray flow in classical chaos in the formation of quasi-bound states in quantum chaos.


A two-dimensional deformed microcavity used in our experiment is made of a liquid jet of ethanol (refractive index $n$=1.361) doped with Rhodamine B dye at a concentration of $10^{-4}$  M/L. The cross sectional shape of the jet is composed of a quadrupole and a much smaller octapole component. Nonethless, our cavity can be approximated by a quadrupole, defined as $\rho(\phi)= r(1+\eta \cos 2\phi)$, with a deformation parameter $\eta$ and a mean radius $r$ of about 15 $\mu$m. The details of our new liquid-jet apparatus is described elsewhere \cite{prep.liquid-jet.Yang05}. 

\begin{figure}
\includegraphics[width=3.4in]{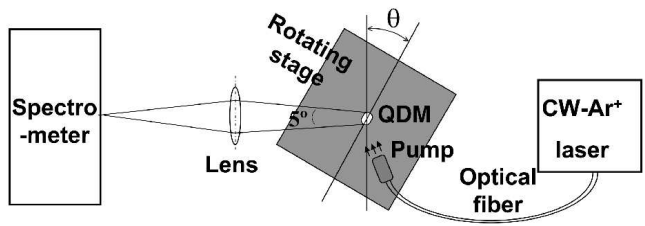}
\caption{Schematic of our experimental setup for measuring cavity mode spectrum and single-mode far-field emission pattern of a QDM.}
\label{fig1}
\end{figure}

As illustrated in Fig.\ \ref{fig1}, the microjet apparatus is mounted in the center of a rotatable stage and the microjet QDM is illuminated by a cw  Ar$^+$ pump laser ($\lambda_p =$514 nm) with its polarization parallel to the QDM column. Since the off-resonance pumping efficiency for a QDM depends on pumping angle \cite{prep.PumpDependence.Lee05}, the pump laser beam is delivered through an optical fiber with its exit end mounted on the rotatable stage with a fixed angle of 45$^\circ$ with respect to the major axis of QDM when output spectrum and directionality are measured. The cavity-modified fluorescence (CMF) or lasing light from the QDM is collected by an objective lens with a full collection angle of 5 degrees and focused on to an entrance slit of a spectrometer. A polarizer placed in front of the slit selects only the polarization component parallel to the QDM column. The emission spectrum is then measured for a fixed angle $\theta$ of the rotation stage.

\begin{figure}
\includegraphics[width=3.4in]{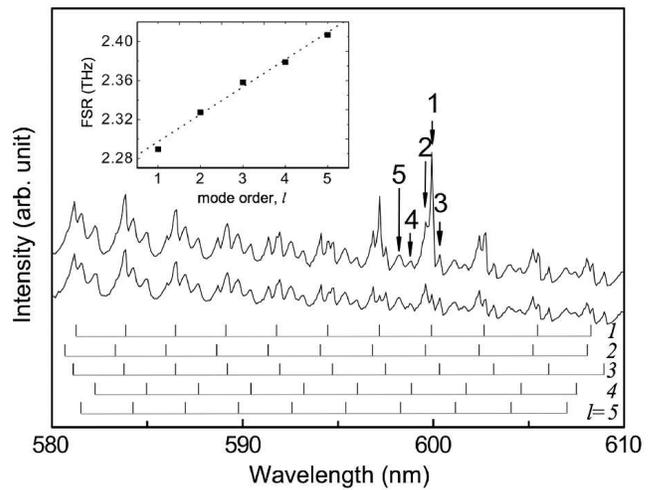}
\caption{Cavity-modified fluorescence and lasing spectra of our QDM with $\eta$=16\% for the pump power of 54 mW  and 112 mW. Inset: Five distinct mode groups are identified and labeled as $l=1, 2, 3, 4, 5$ with increasing order of FSR. The positions of these mode groups are marked as vertical ticks below the spectra.}
\label{fig2}
\end{figure}

Figure \ref{fig2} shows the resulting CMF and lasing spectra of a QDM with $\eta$=16\%. Detection angle $\theta$ was fixed at $\theta$=40$^\circ$ for various pumping powers. We can identify five distinct mode groups, each of which consists of a sequence of peaks with a well-defined FSR. Modes in the same mode group differ from each other by the same FSR in frequencies and share similar modal properties such as quality factor and spatial distribution. By examining their FSR values, we can {\em uniquely} label these mode groups as $l=1, 2, 3, 4, 5$ in increasing order of FSR in frequency, as shown in the inset of Fig.\ 2, in analogy to radial mode order for a circular cavity. Some mode groups undergo laser oscillations as the pumping power increases.

Since the instrumental resolution of spectrometer (0.2 nm) limits the largest $Q$ value to be directly observed to $\sim 10^3$, we employed two independent methods to measure $Q$ values of the observed mode groups. First, in the CMF spectra as shown in Fig.\ \ref{fig2} the $Q$  value is determined from the  wavelength at which the visibility of a mode drops as we scan from long to short wavelengths since at that wavelength the cavity loss  approximately equals the known absorption loss of the cavity medium \cite{Chylek91}. Secondly, the $Q$  value can also be measured from the wavelength at which a mode undergoes laser oscillation at threshold as seen in Fig.\ \ref{fig2} \cite{Moon00,Lee02}. Both methods consistently give $Q = 5\times 10^5$ for $l=1$. For $l=2, 3,4,5$ only the CMF method is used since their $Q$'s are not large enough to undergo laser oscillation at the prefixed dye concentration. The results are $Q =2\times 10^5$ for $l=2$, $Q=2\times 10^4$ for $l$=3 and 4 and $Q = 1\times 10^4$ for $l=5$.

\begin{figure}
\includegraphics[width=3.4in]{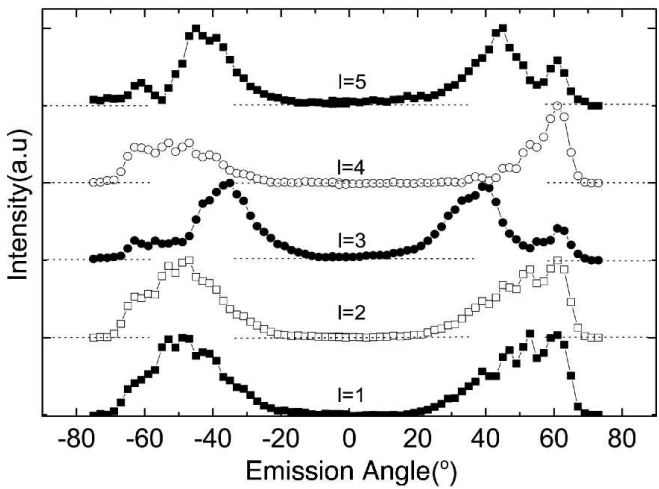}
\caption{Observed far-field emission pattern of individual single cavity modes. The observed single modes are labeled from bottom as $l=1, 2, 3, 4, 5$. Baseline for each curve is denoted by dotted lines.}
\label{fig3}
\end{figure}

Output emission directionality of each {\em single} mode can be measured by taking spectra like Fig.\ \ref{fig2} for various rotation angles $\theta$ with respect to the direction of minor axis of the QDM and then plotting the peak height of each mode as a function of $\theta$, which also equals the far-field emission angle in Figs.\ 3 and 5. By this method, contribution by background luminescence, mostly due to bulk fluorescence of the cavity medium, can be discriminated. The resulting single-mode far-field emission pattern is shown in Fig.\ \ref{fig3}.
In comparison, previously reported far-field patterns for deformed microcavities of $nkr\sim 10^3$ \cite{Schwefel04} were measured without any resolution on individual modes and thus they correspond to multimode far-field emission patterns.

The most striking feature of the far-field patterns in Fig.\ \ref{fig3} is that they all look similar, highly directional and peaked around 40$^\circ$-60$^\circ$. One plausible explanation for the observed similarity would be that all five mode groups might have very similar intracavity mode distributions regardless of their completely different $Q$ values. However, this is quite unlikely since different $Q$ values by many orders of magnitude are usually associated with significantly different mode distributions. In fact, we show below theoretically that these mode groups indeed have quite different mode distributions inside the cavity.

\begin{figure}
\includegraphics[width=3.4in]{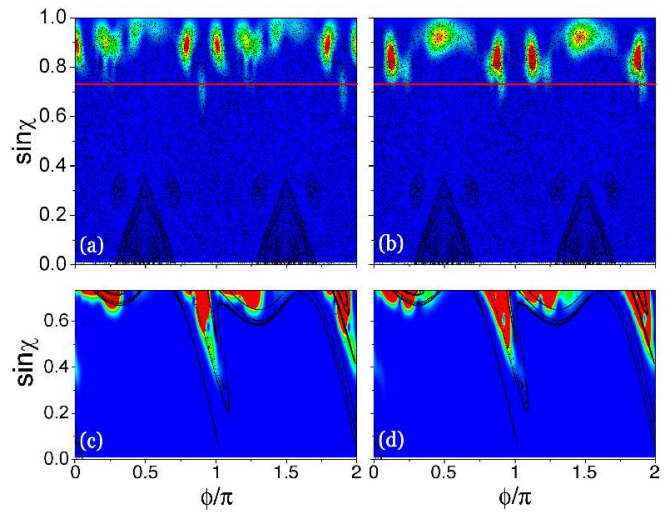}
\caption{Calculated Husimi Plots of quasieigen modes with (a) $nkr=$ 232.27 and (b) $nkr=$ 226.29, corresponding to  $l$=2 and 4 modes, respectively. Distributions below critical line ($=1/n$) in (a) and (b) are enhanced 100 times in intenisty in (c) and (d), respectively.}
\label{fig4}
\end{figure}

Consideration of the ray dynamics in a QDM in classical limit ($nkr\rightarrow \infty$) often sheds light on mode formation in a QDM of a finite $nkr$. The ray dynamics can be completely analyzed in terms of a Poincar\'e surface of section (PSOS) obtained by recording both the angle of incidence $\chi$ and the azimuthal angle $\phi$ for successive reflections off the cavity boundary. PSOS's for a quadrupole with $\eta=0.16$, displayed as backdrops in Figs.\ \ref{fig4}(a) and 4(b), show that there exist no stable periodic orbits above the line of critical angle, $\sin \chi_c=1/n$. None of five mode groups observed in our experiment can thus be associated with stable periodic orbits.

Modes or quasi-bound state solutions of Maxwell's equations for an open QDM with $\eta$=16\% and $n$=1.361 are calculated by using the boundary element method \cite{Kagami84,Wiersig03}. For direct comparison with experiment, we search for modes in the same range of $nkr$ as in the actual experiments. In the eigenvalues of the quasi-bound states numerically obtained, we also observe five distinct mode groups with well-defined mode spacing or FSR, according to which modes are labeled as $l=1, 2, 3, 4, 5$ as before. Their $Q$ values are comparable to those of experiment.

\begin{figure}
\includegraphics[width=3.4in]{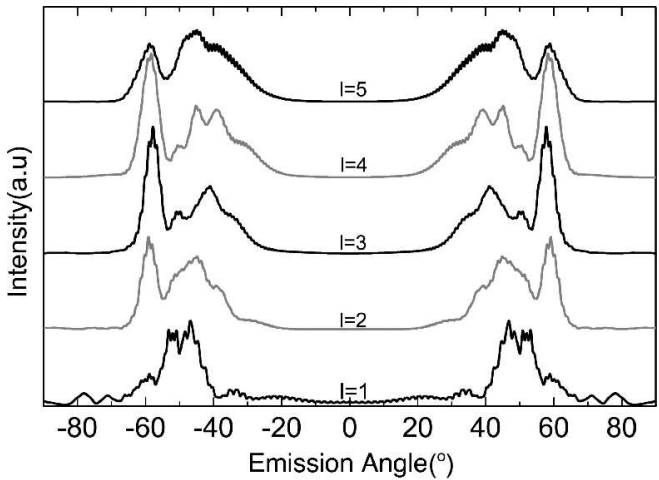}
\caption{Calculated far-filed distributions associated with quasieigenmodes with $nkr$=228.87, 232.27, 226.29, 227.65 and 230.87, which belong to mode groups labeled as $l=1, 2, 3, 4, 5$, respectively. Curves are smoothed over $5 ^\circ$ intervals to match the experimental angular resolution.}
\label{fig5}
\end{figure}

Husimi distributions of some of the modes ($l=2,4$) are shown in Fig.\ \ref{fig4}, superimposed on the classical PSOS. The Husimi distribution is a Gaussian-smoothed version of the Wigner function, representing the corresponding quantum mechanical probability distribution in phase space \cite{Reichl92,Crespi93}. We confirm that modes belonging to the same mode group show almost the same shape of wavefunction \cite{prep.PRA.Shim05}. Some of the mode groups are localized on the UPO's, and thus categorized as scar modes: we can associate an octagon with $l=2$ mode and  a hexagon with $l=4$ mode in Fig.\ \ref{fig4}. The five mode groups exhibit qualitatively different spatial distributions, barely overlapping in Husimi plot. Complete results on mode calculations and analyses on our experiment can be found in Ref.\ \cite{prep.PRA.Shim05}.
The far-field emission pattern can be calculated once a quasi-bound state is known. Interestingly, all five mode groups give rise to very similar far-field patterns as shown in Fig.\ \ref{fig5}, supporting the experimental observation
in Fig.\ \ref{fig3}.


So far, it has been widely accepted that the modes with distinct shapes exhibit different output directionality \cite{Gmachl02,Harayama03,Rex02,sofar}, so the above experimental and theoretical results contradict to the conventional wisdom. This seeming contradiction is, however, resolved by close examination on the phase-space mode distribution in the Husimi plot. When we enhance the intensity near the line of critical angle by 100 times, we can see a few tail-like faint structures extending from the regions of wave concentration to the region far below the line of critical angle. Integrated probability associated with these structures below the line of critical angle is negligibly small compared to that of the total mode distribution for $l=2, 4$ in Fig.\ \ref{fig4}. No matter how small their probability is, they are the only direct routes to outside world by refraction. Without them, light would have to tunnel through a substantial distance in phase space from the region of wave concentration to outside across the line of critical angle, and thus the probability of such tunneling would be extremely low. For the other mode groups similar faint structures also exist, providing routes for refractive escape of light.

Interestingly, the origin of these faint route structures in the wave function can be traced back to the classical ray chaos. A hint comes from the observation that the far-field emission pattern obtained from ray dynamics in classical limit for the QDM is similar to those from the wave calculation and the experiment. For an open system, long time ray dynamics in chaotic region are predominantly determined by the so-called unstable manifolds \cite{Lichtenberg83,Reichl92}, which are drawn in Fig.\ \ref{fig4}(c) and (d) in dark lines for our QDM. Since rays escapes from an open cavity before reaching completely ergodic limit, ray dynamics is usually restricted in limited phase space and thus follows a few dominant unstable manifolds.

This scenario is, however, valid in the classical limit $nkr\rightarrow \infty$, and the long time dynamics in this picture usually corresponds to several ray circulations, which is much shorter than $10^{2-3}$ ray circulations corresponding to $Q\sim 10^5 - 10^6$ observed in our experiment. Nevertheless, the underlying structure of unstable manifold imposed by the geometric shape of the deformed cavity persists as a weak quantum localization, with $1/nkr\sim1/230$ playing a role of an effective Planck constant. Although the quantum localization around the unstable manifold is not as strong as that for UPO's in the formation of scar modes, it leaves a trace as a few faint structures in phase-space mode distribution across the line of critical angle, resulting in the observed directional output.

It is noted that for a mode with rather low $Q$ the Husimi distribution has significant overlap with the region below the critical angle and thus the
output directionality is mainly determined by the intracavity mode distribution as usual \cite{prep.PRA.Shim05}. In addition, for high-$Q$ modes for rather small $nkr$, namely $\sim 50$, which happens to be the size parameter at which many other theoretical studies have been performed on the role of unstable manifolds in output directionality \cite{Schwefel04}, the faint structure corresponding to the unstable manifolds has not been observed in our numerical studies \cite{prep.PRA.Shim05}.

Our results on universal output directionality can greatly simplify design considerations for microcavity devices; a highly directional output can be achieved regardless of intracavity modal distributions and $Q$ values of individual modes in significantly deformed microcavities.

This work was supported by National Research Laboratory Grant and by KRF Grant (2005-070-C00058). JBS and HWL were supported by a Grant from the Ministry of Science and Technology of Korea. SWK was supported by KRF Grant (2004-005-C00044) and by KOSEF Grant (R01-2005-000-10678-0).


\begin{thebibliography}{99}

\bibitem{Q-chaos}
H.-J.\ Stoeckmann, ''Quantum Chaos: an introduction'' (Cambridge Univ.\ Press, Cambridge, 1999).

\bibitem{Anderson}
P.\ A.\ Lee and T.\ V.\ Ramakarishnan, Rev.\ Mod.\ Phys.\ {\bf 57}, 297 (1985).

\bibitem{Lagendijk}
D.\ S.\ Wiersma, P.\ Bartolini, A.\ Lagendijk, and R.\ Righini, Nature {\bf 390}, 671 (1997).

\bibitem{Genack}
A.\ A.\ Chabanov, M.\ Stoytchev, and A.\ Z.\ Genack, Nature {\bf 404},850 (2000). 

\bibitem{hydrogen}
J.\ E.\ Bayfield, G.\ Casati, I.\ Guarneri, and D.\ W.\ Sokol, Phys.\ Rev.\ Lett.\ {\bf 63}, 364 (1989).

\bibitem{quantumdot}
C.\ M.\ Marcus {\em et al.},
Phys.\ Rev.\ Lett.\ {\bf 69}, 506 (1992).

\bibitem{Sridhar}
S.\ Sridhar, Phys.\ Rev.\ Lett.\ {\bf 67}, 785 (1991).

\bibitem{photonics}
K.\ Vahala, Nature {\bf 424},839 (2003).

\bibitem{Noeckel}
J.\ U.\ Noeckel and A.\ D.\ Stone, Nature {\bf 385}, 45 (1997).

\bibitem{Reichl92}
L.\ E. Reichl, {\em The Transition to Chaos in Conservative Classical Systems: Quantum Manifestations} (Springer-Verlag, New York, 1992).

\bibitem{Lichtenberg83}
A.\ J.\ Lichtenberg and M.\ A.\ Lieberman, {\it Regular and Stochastic Motion} (Springer-Verlag, New York, 1983).

\bibitem{Gmachl02}
C.\ Gmachl {\em et al.},
Opt. Lett. {\bf 27}, 824 (2002).

\bibitem{Harayama03}
T.\ Harayama {\em et al.},
Phys. Rev. E {\bf 67} 015207 (2003).

\bibitem{Schwefel04}
H.\ G.\ Schwefel {\em et al.},
J.\ OPt.\ Soc.\ Am.\ B {\bf 21}, 923 (2004).

\bibitem{Rex02}
N.\ B.\ Rex {\em et al.},
Phys. Rev. Lett. {\bf 88}, 094102 (2002).

\bibitem{Lee02}
S.-B .\ Lee {\em et al.},
Phys. Rev. Lett. {\bf 88}, 033903 (2002).

\bibitem{prep.liquid-jet.Yang05}
J.\ Yang {\em et al.}, Rev.\ Sci.\ Instrum.\, {\bf 77}, 083103 (2006).

\bibitem{prep.PumpDependence.Lee05}
S.-B.\ Lee {\em et al.}, 
arxiv.org/abs/physics/0603158.

\bibitem{Chylek91}
P.\ Chylek {\em et al.},
Opt. Lett. {\bf 22}, 1723(1991).

\bibitem{Moon00}
H.-J.\ Moon, Y-T.\ Chough, and K.\ An, Phys. Rev. Lett. {\bf 85}, 3161 (2000).

\bibitem{Kagami84}
S.\ Kagami and I.\ Fukai, IEEE Trans. Microwave Theory and Techniques
{\bf 32} 455 (1984).

\bibitem{Wiersig03}
J.\ Wiersig, J.\ Opt.\ A {\bf 5} 53 (2003).

\bibitem{sofar}
Scott Lacey{\em et al.},
 Phys. Rev. Lett. {\bf 91}, 033902 (2003).

\bibitem{Crespi93}
B.\ Crespi, G.\ Perez, and S.-J.\ Chang, Phys. Rev. E {\bf 47}, 986 (1993).

\bibitem{prep.PRA.Shim05}
J.-B.\ Shim {\em et al.}, arxiv.org/abs/physics/0603221.


\end{thebibliography}
\end{document}